# Domain Structures in Confined Nanoferroelectrics: Phase Field Approach

Julia Slutsker[1], Andrei Artemev[2] and Alexander Roytburd[3]

[1]Ceramics Division, MSEL, NIST, Gaithersburg, 20899, USA

[2]Department of Mechanical and Aerospace Engineering, Carleton University, 1125 Colonel By Drive, Ottawa, ON, K1S 5B6 Canada

[3]Department of Materials Science and Engineering, University of Maryland, College Park, MD 20742, USA

## Abstract

Phase field modeling of domain structures in ferroelectrics nanorods of different shape and sizes is presented. The vortex domain configurations in confined ferroelectrics have been explored by varying the ratio of the energies of electrostatic and elastic interactions. It is shown that a strong effect of the electrostatic interactions can cause the formation of $90^o$ domain walls that do not satisfy the condition of strain compatibility. A good agreement between the results of phase field modeling and the results of atomistic calculations for nano ferroelectrics demonstrates that the phase field approach provides an effective tool for the analysis of domain structures in nano-ferroelectrics.

The interest in the nanoferroelectrics has increased in the last decade [1-4]. Small ferroelectric structures such as disks, rods, wires and tubes are promising candidates for the ferroelectric memory as well as for other applications requiring miniaturizing, including transducers and actuators, multifunctional sensors, and tunable microwave devices [5-7].

Although, there are no experimental data on domain structures in confined nanoferroelectrics, the first-principles-based simulations have provided an important insight into domain structures in nanodimension ferroelectrics. These simulations have revealed a significant effect of the depolarizing field on dipole patterns of polarization in low dimensional ferroelectrics. It has been shown that in low dimensional ferroelectrics such as disks, rods, and wires the screening of the depolarizing field proceeds through the alignment of polarization along the surface. It results in the formation of vortex dipole structures in ferroelectric nanoparticles [8-12]. These structures minimize the electrostatic energy of the depolarizing field in the same way as closed flux domain configurations minimize the magnetostatic stray field in ferromagnetics [13].

The results of first-principles-based simulations are essential for understanding physics of nanoferroelectrics and for stimulating experimental studies in this area. This method provides quantitative characteristics of ferroelectrics for very small particles. However, the analysis of larger volumes suitable for experimental observations requires a considerable increase in computational efforts. Therefore, the simulations based on the phase field approach appear to be an effective complimentary tool for the investigation of domain structures in nanoferroelectrics on a larger scale including a mesocopical one. On the other side, since the characteristic length (i.e., the domain wall thickness) in ferroelectrics is about 1 nm, the phase field approach provides an adequate description of domain structures in ferroelectric nanoparticles down to tens of nanometers [14]. The phase field modeling has proven to be a successful method for describing and predicting domain structures and functional properties of ferroelectric films of a sub-micron size [15-19]. However, all the previous phase field modeling of domain structures in ferroelectrics has been performed for thin films with either a fully compensated depolarizing field (films under short circuited conditions) [15-17] or significantly diminished electrostatic interactions due to the introduction of the dielectric constant of ferroelectrics into the energy of dipole-dipole interactions [18,19]. Thus, the simulated domain structures in thin films have been mainly dictated by elastic interactions. Strong electrostatic interactions together with elastic interactions are taken into account in the present work.

In this letter we present the results of a study of the domain structure in confined ferroelectrics. We use an important advantage of the phase field approach which allows us to vary independently the parameters determining domain structures and, therefore, we can investigate a wide spectrum of possible domain configurations. Particularly, we consider the effect of the ratio of the energies of electrostatic and elastic interactions in nanoferroelectrics on their domain structures. The method presented here allows one to simulate domain structures in confined ferroelectrics of different shapes and sizes under different electrical and mechanical boundary conditions as well as the effect of external electrical and mechanical fields.

We are focusing on confined ferroelectric crystals that are elements of a nanocomposite. As an example, we consider a periodic nanostructure containing a ferroelectric nanorod embedded into a non-polarized matrix (Fig.1). Such kind of nanostructure has been developed by the epitaxial self-assembling of lattice-matched but immiscible phases on single-crystal substrates [20-24].

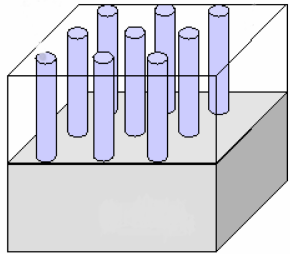

*Fig. 1. Periodic nanostructure containing ferroelectric rods embedded into a non-polarized matrix.*

An equilibrium distribution of polarization corresponding to the equilibrium domain structure in the ferroelectric rods corresponds to the minimum of the free energy functional:

$$F = F_{electro} + F_{el} + F_{GL} \qquad (1)$$

where $F_{electro}$ is the energy of electrostatic interactions, $F_{el}$ is the energy of elastic interactions, and $F_{GL}$ is the Ginzburg-Landau potential. The following represents the energy of long-range electrostatic dipole-dipole interactions:

$$F_{electr} = \frac{1}{8\pi \cdot \varepsilon_0 \varepsilon^*} \iint_V \left[ P_i(\mathbf{r}) \left( \frac{\delta_{ij}}{|\mathbf{r}-\mathbf{r}'|^3} - \frac{3(r_i - r_i')(r_j - r_j')}{|\mathbf{r}-\mathbf{r}'|^5} \right) P_j(\mathbf{r}') \right] dV - E_i^{appl} \int_V P_i(\mathbf{r}) dV \qquad (2)$$

where $P_i(r)$ is a polarization, $E_i^{appl}$ is an external electric field, $\varepsilon_0$ is a dielectric constant of vacuum, and $\varepsilon^*$ is a high frequency dielectric constant is not associated with the ferroelectric order parameter, $\varepsilon^*$ is in order of 1 [11, 25]. To emphasize the effect of electrostatic interactions in our simulations we assume that $\varepsilon^* = 1$, contrary to the most phase field simulations of a ferroelectric domain structure where $\varepsilon^*$ has been assumed as the dielectric constant of ferroelectrics [18,19]. The energy of long-range elastic interactions is determined as

$$F_{el} = \int_V \left[ \frac{1}{2} C_{ijkl} (\varepsilon_{ij}(\mathbf{r}) - \varepsilon_{ij}^0(\mathbf{r}))(\varepsilon_{kl}(\mathbf{r}) - \varepsilon_{kl}^0(\mathbf{r})) dV \right] - \sigma_{ij}^{appl} \int_V \varepsilon_{ij}^0(\mathbf{r}) dV \qquad (3)$$

where $C_{ijkl}$ is the elastic modulus, $\varepsilon_{ij}(r)$ is the total strain, and $\sigma_{ij}^{appl}$ is an external stress. The transformation self-strain, $\varepsilon_{ij}^0(\mathbf{r}) = Q_{ijkl} P_i(r) P_k(r)$, where $Q_{ijkl}$ is an electrostrictive coefficient tensor. The elastic energy associated with a ferroelectric nanorod embedded in a composite consists of two parts: the energy of a non-uniform distribution of polarization and the energy of elastic interactions of a nanorod with a matrix. The elastic moduli are assumed to be equal in the rod and a matrix.

The distribution of polarization $P_i(r)$ is characterized by the field of the order parameter $\eta_i(r)$, $P_i(\mathbf{r}) = P^0 \cdot \eta_i(\mathbf{r})$, where $P^0$ is the saturation polarization of a single domain equilibrium state. The energy of the order parameter field is described by Giznsurg-Landau potential, $F_{GL}$:

$$F_{GL} = \int_V \left[ f_0(\eta_i) + \frac{1}{2} \beta_{ijkl} \frac{\partial \eta_i}{\partial x_k} \frac{\partial \eta_j}{\partial x_l} \right] dV, \qquad (4)$$

where $f_0(\eta_i)$ is the free energy density described by the Landau-Devonshire type expansion for the three component order parameter, $\eta_i$, $i=1,2,3$:

$$f_0(\eta_i) = \alpha_1(\eta_1^2 + \eta_2^2 + \eta_3^2) + \alpha_{11}(\eta_1^4 + \eta_2^4 + \eta_3^4) + \alpha_{12}(\eta_1^2\eta_2^2 + \eta_2^2\eta_3^2 + \eta_3^2\eta_1^2) + \\ \alpha_{111}(\eta_1^6 + \eta_2^6 + \eta_3^6) + \alpha_{112}(\eta_1^2(\eta_2^4 + \eta_3^4) + \eta_2^2(\eta_1^4 + \eta_3^4) + \eta_3^2(\eta_1^4 + \eta_2^4)) + \alpha_{123}\eta_1^2\eta_2^2\eta_3^2 \quad (5)$$

This potential with coefficients corresponding to the transformation from cubic to tetragonal states in PbTiO3 is used in the simulations [26]. The second term in Eq.(4) where $\beta_{ijkl}$ is the gradient coefficient tensor expresses short-range interactions in ferroelectrics and determines the interface energies. The value of the components of the gradient coefficient tensor, $\beta_{ijkl}$, cannot be determined with a feasible accuracy due to the difficulties with the separation of short and long range interactions. Therefore, the estimation of the anisotropy of this coefficient can be ambiguous. In our simulations we neglect this anisotropy as it has been done in other work on phase field modeling of ferroelectrics [15-19]. By neglecting anisotropy of the gradient coefficient tensor we simplify the structure and energy of interfaces. Our preliminary simulations with the anisotropic gradient coefficient tensor did not qualitatively change the equilibrium domain structures. Therefore, it is possible to conclude that the domain structure does not considerably depend on the details of the domain wall structure if the domain sizes are much large than the domain wall thickness. It is worth noting that this assumption does not exclude the orientation dependence of domain interfaces due to the effect of long-range interactions on the intrinsic interface energies [27]. This problem needs clarification through a careful comparison with the results of first-principle calculations.

The equilibrium domain structure that corresponds to the equilibrium field of order parameters $\eta_i^0(\mathbf{r})$ is a solution of the equation

$$\frac{\delta F}{\delta \eta_i(\mathbf{r})} = 0 \quad (6)$$

where $\delta/\delta\eta(\mathbf{r})$ is a variational derivative. This equilibrium field can be obtained by solving the time-dependent Ginzburg-Landau equation with a fast Fourier transform implementation for the energy of the dipole-dipole interaction [28]:

$$\frac{\partial \eta_i}{\partial t} = -L\frac{\delta F}{\delta \eta_i} + \xi_i \quad (7)$$

where $L$ is a kinetic coefficient and $\xi_i$ is the Langevin noise term. According to the microelasticity approach in the phase field modeling [28-31], the evolution of the microstructure is described by the solution of Eq.(7) presented in the reciprocal space. The following dimensionless parameters have been used in the simulations: dimensionless Landau coefficients, $\alpha/\Delta f$, where $\Delta f$ is the difference between paraelectric ($\eta_i = 0$) and ferroelectric states ($\eta_i = \eta_i^0$) ($\frac{\partial f_0(\eta_i)}{\partial \eta_i} = 0$). The dimensionless gradient coefficient is $\tilde{\beta} = \frac{\beta}{\Delta f l_0^2}$, where $l_0$ is the computation grid length. The gradient coefficient, $\tilde{\beta}$, determines the calculation length scale, $l_0 = l\sqrt{\frac{1}{\tilde{\beta}}}$, where $l$ is the thickness of ferroelectric domain walls ($l \sim 1\text{nm}$).

The characteristic elastic energy is $\zeta = \left(\frac{\bar{\varepsilon}_{ij}^0 \cdot C_{ijkl} \cdot \bar{\varepsilon}_{kl}^0}{2}\right)/\Delta f$, where $\bar{\varepsilon}_{ij}^0 = Q_{ijkl} P_i^0 P_k^0$. The characteristic energy of electrostatic interactions is $\lambda = \left(\frac{P^{0^2}}{2\pi\varepsilon_0}\right)/\Delta f$.

A 512x64x64 mesh with periodic boundary conditions is used for the simulation of a rod surrounded by a dielectric matrix (a symmetrical cell, Fig.2a). The electrostatic image charge principle was used so that the half of this cell, 256x64x64, represented the rod with one of its ends covered by an electrode (an assymetrical cell, Fig.2b).

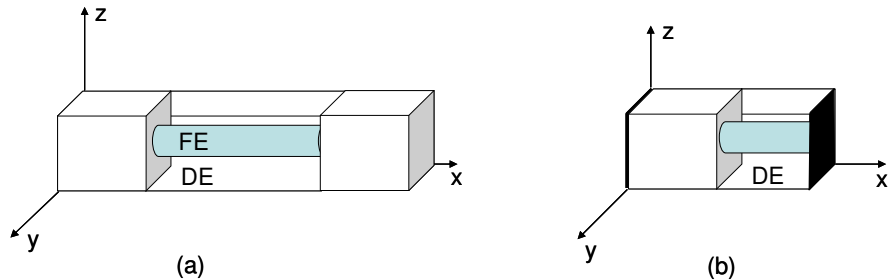

(a)      (b)

*Fig. 2. (a) Computational cell containing a ferroelectric rod embedded into a non-polarized matrix; (b) The half of the computational cell representing the rod with one of its ends covered by an electrode.*

The transformation from an unstable paraelectric occurs through the nucleation of a stable ferroelectric phase driven by the Langevin noise in the evolution equations (7). After a sufficiently long relaxation process the equilibrium is established between all domains and in the final part of the simulation process the domain pattern remains unchanging.

Domain structures have been investigated in confined ferroelectric nanorods with different cross-section shapes and sizes. To reveal relative effects of electrostatic and elastic energies two different $\lambda/\zeta$ ratios have been used in the simulations: $\lambda/\zeta=25$ corresponding to ferroelectrics with a large spontaneous deformation, such as PbTiO3 or BaTiO3 at room temperature, and $\lambda/\zeta=250$, corresponding to ferroelectrics with a small spontaneous deformation. A similar increase in value of $\lambda/\zeta$ can be the result of the decreasing saturation polarization in ferroelectrics with an increasing temperature.

Fig.3 presents domain structures in a rod along <100> with a square cross-section and {100} faces. The modeling has been performed with the gradient coefficient $\tilde{\beta}=8$ corresponding to the cross-section length of 16 nm and with the ratio of elastic to electrostatic energies, $\lambda/\zeta$, equals to 25. The domain structure presents closed circuits of 90$^o$ domains (Fig.3a). The presence of the triangle 90$^o$ domain at the end of the rod completely screens the electrostatic stray field. It should be emphasized that the presented structure is an equilibrium one: it can be removed by an applied electrical field and can appear again after the field is removed. The cubic 90$^o$ domains in the middle of the constrained rod enable the maximal relaxation of the energy of elastic interactions between a rod and a matrix. An increasing relative contribution of the elastic energy leads to the increase in the number of cubic domains, while these domains disappear when the elastic energy decreases its contribution (Fig.3b, $\lambda/\zeta=250$). The equilibrium distance between the cubic domains is determined by the competition between the increasing energy of the disclination distorsions around the edges of these domains and the increasing constraint energy. If the aspect ratio of the nanorod decreases toward the cubic shape, the vortex-like domain structures similar to ones obtained by the first-principles-based simulations are formed [10,11].

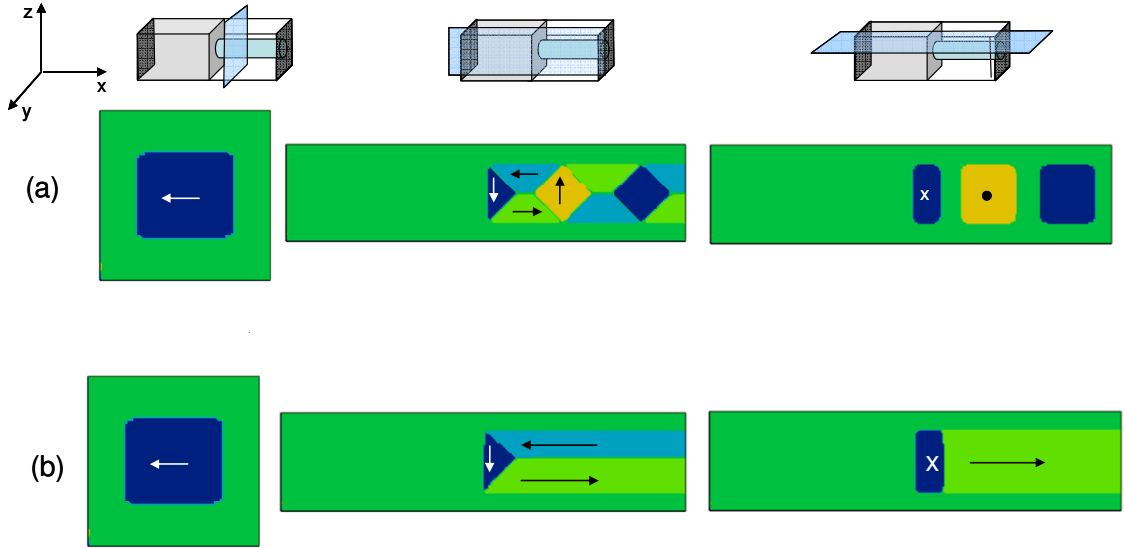

*Fig. 3. Domain structure in square shaped rods with the cross-sectrion length of 16 nm (half of the computational cell are shown). (a) Ratio of electrostatic and elastic energies, $\lambda/\xi$=25. (b) Ratio of electrostatic and elastic energies, $\lambda/\xi$=250.*

The simplest structure with one closure domain obtained in the system with a small contribution of the elastic energy (Fig.3b) becomes more complex if the dimensions of the rod cross-section increase. As an example, we present the results of the simulations of domain structures at $\lambda/\zeta$ =250 and $\tilde{\beta}$ =2 corresponding to the cross-section length of 35nm (Fig.4). In this case the vortex configuration of four 90$^o$ domains that completely screens the electrostatic stray field at the end of the rod is forming. A cylindrical configuration of 180$^o$ domains is formed far from the end of the rod (the introduction of the anisotropy of the gradient energy results in the faceting of the cylindrical domains). These two structures are connected through the system of plane 90$^o$ domain walls oriented along {111} planes (Fig.4b). It should be noted that although the domain walls are oriented along the {111} plane, they are still free of charge. The condition for the electro-neutral domain boundary, $\Delta\mathbf{P}\cdot\mathbf{n}$=0, leaves one of **n** components free and, therefore, allows for the {111} oriented domain boundaries and even for curved 90$^o$ domain boundaries without charge.

The domain structures in nanorods with circle cross-sections of different diameters are presented in Figs. 5 and 6. Figure 5 presents the domain structure in a nanorod at $\tilde{\beta}=8$ corresponding to the diameter of 16nm and with $\lambda/\xi$=25. This domain structure is similar to the one obtained in the square rod (Fig. 4a), however, unlike the square shaped rods the domain walls between 90° domains are cylindrical surfaces (Fig.5b). Although there are no electrical charges on these curved domain walls, they are the sources of internal stresses. Their presence in the equilibrium domain structures is dictated by the shape of rods and is a result of a relatively small effect of elastic energy in comparison to the electrostatic energy of the depolarizing field.

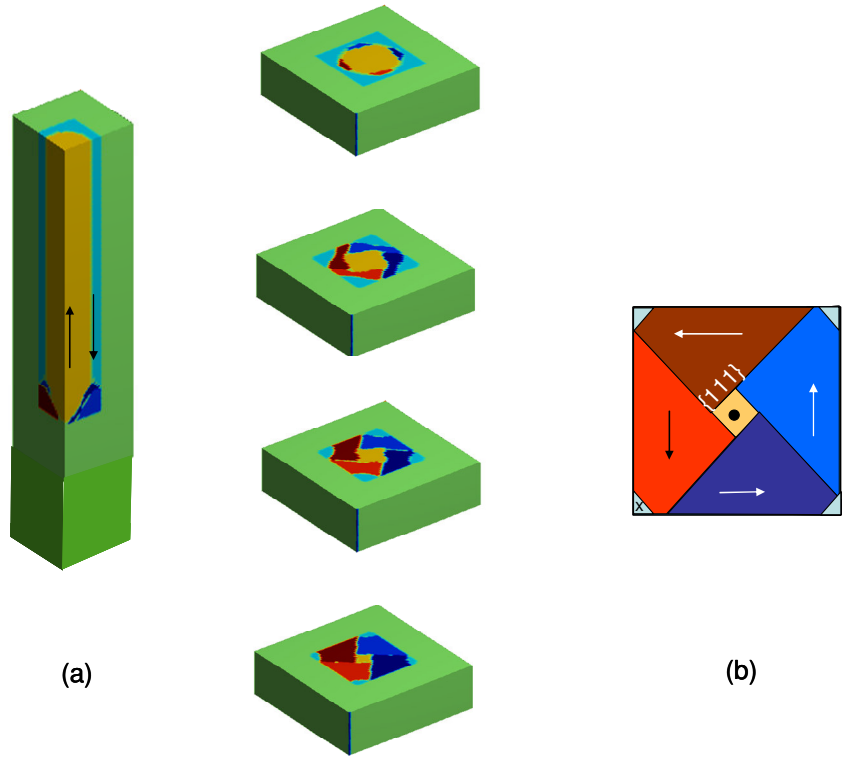

*Fig. 4. Domain structure in the square shaped rod with the cross-section length of 35 nm and ratio of electrostatic and elastic energies, $\lambda/\xi$=250. (a)Series of plane sections along the ferroelectric rod, from the end of the rod (bottom) toward the middle of the rod (top). (b) Schematic picture of domain walls in a ferroelectric rod.*

The complexity of the domain structure in the circle shaped nanorods increases with the increase in the rod diameter. Figure 6 presents the domain structures in a nanorod with $\tilde{\beta}=2$ corresponding to the diameter of 35 nm and with $\lambda/\xi$=25. The stray field at the end of the rod is compensated by the formation of the vortex configuration consisting of four 90° domains with domain walls oriented along {111}. The cylindrical 180° domains are formed far from the end of the rod. The vortex-like domain configuration with in-plane polarization is connected with an outer out-of-plane domain along the {111} domain walls and is connected with an inner out-of-plane domain along the distorted {110} domain walls. This distorsion of the {110} plane walls as well as the curved 90° domain walls in Fig.5 is a result of an uncompensated effect of the depolarizing field from the cylindrical surface of the nanorod. The distribution of polarization inside domains in circle shaped rods is less uniform than the distribution of polarization in square shaped rods. A similar nonuniform polarization distribution and curved domain walls have been obtained by the first-principle-based simulations of ferroelectric spherical nanodots [10,11].

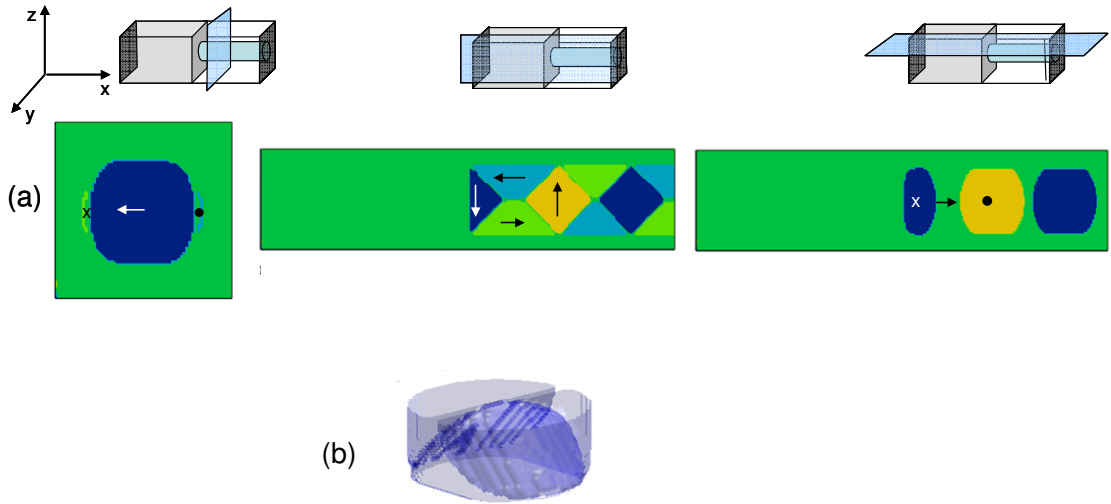

Fig. 5. *Domain structure in a circle shaped rod of a diameter of 16 nm and the ratio of electrostatic and elastic energies, $\lambda/\xi=25$. (a) Domain structure in y-z, x-z and x-y planes. (b) Domain boundaries of 90º domains inside the ferroelectric rod.*

The structures presented above are expected to be found in confined ferroelectrics where the effect of electrostatic interactions, particularly, the depolarizing field is not diminished by the screening charges. Since the formation of 180° domains is not effective for the screening of the depolarizing field in small ferroelectric particles, this field is reduced by the formation of the circuits of the 90° domains with a triple or four-fold junction. As shown in this letter a strong effect of the electrostatic interactions can cause the formation of 90° domain walls that do not satisfy the condition of strain compatibility.

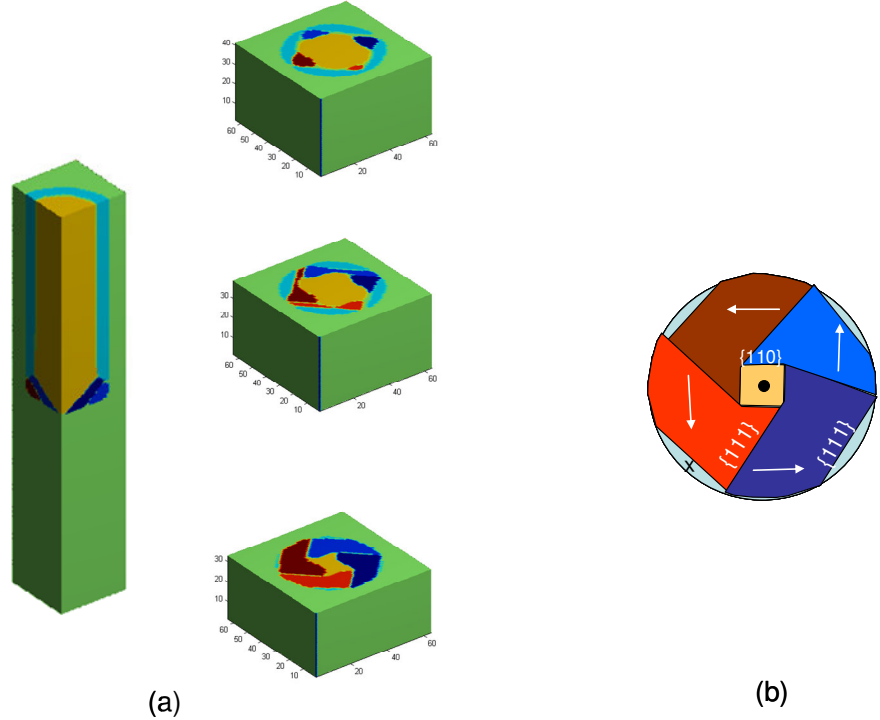

*Fig. 6. (a) Series of plane sections along the circle shaped ferroelectric rod with the diameter of 35 nm and ratio of electrostatic and elastic energies, $\lambda/\xi=25$. Sections show domain structure from near the end of the ferroelectric rod (bottom) toward to the middle of the rod (top). (b) Schematic picture of the domain walls in a ferroelectric rod.*

These domain walls are charge free, but they can have an unusual plane orientation or be curved. Similar domain structures are observed in ferromagnetic rods and particles [13] wherein elastic interactions can be neglected. However, the effect of elastic interactions cannot be neglected in ferroelectrics. Since the spontaneous strain in ferroelectrics is much larger than magnetostriction in ferromagnetics, the formation of closure domains in ferroelectrics results in a strong disclination distorsion accompanied by a large elastic energy. Therefore, vortex-like configurations can be the equilibrium elements of domain structures only in nanosize ferroelectrics where the formation of closed domain circuits is dictated by the trend to screen the depolarizing field and minimize the electrostatic energy which plays a dominant role in the thermodynamics of confined ferroelectrics.

A good agreement between the results of the phase field modeling obtained for small size nanorods and the results of atomistic calculations obtained for nano ferroelectrics of a similar size and shape demonstrates that the phase field approach provides an effective tool for the analysis of domain structures in nano-ferroelectrics. This approach can be used to study more complex domain patterns appearing in nano-ferroelectric composites with larger scales that make the use of atomistic simulations too computationally expensive. The phase field model can also be applied to the study of the effect of physical properties on the domain structure for a broad spectrum of parameter values.

**Acknowledgments**

A.A. gratefully acknowledges the support of NSERC under the grant RGP 155157-2002. A.R is grateful to the financial support of NSF and Israel-USA Binational Science Foundation.